\documentclass[preprint,amsfonts,onecolumn,nofootinbib]{revtex4}
\begin{document}

\title{On the Large Scale CMB Polarization}

\author{Paolo  Cea}
\email{paolo.cea@ba.infn.it}

\affiliation{Dipartimento di Fisica, Universit\`{a} di Bari, I-70126 Bari, Italy}
\affiliation{INFN - Sezione di Bari, I-70126 Bari, Italy}

\begin{abstract}
We discuss the large scale  polarization of the cosmic microwave background induced by 
the anisotropy of the  spatial geometry  of our universe. Assuming an
eccentricity at decoupling of about  $0.64 \; 10^{-2}$, we find an average large scale polarization
$\Delta T_{pol}/ T_0  =  (0.5 \;  -  \; 1.0) \;  10^{-6}$. We suggest that the fortcoming polarization data at large scales from Planck will
be able to discriminate between our proposal and  the generally accepted reionization scenario.
\end{abstract}

\maketitle

The latest results from the Wilkinson Microwave Anisotropy Probe (WMAP)~\cite{Komatsu:2009} showed that the cosmic microwave background
(CMB) anisotropy data are in remarkable agreement with the simplest inflation model. At large scale, however, some anomalous
features have been reported. The most important discrepancy resides in the low quadrupole moment, which signals an important
suppression of power at large scales, although the probability of quadrupole being low is not statistically significant. \\
\indent
Recently~\cite{Campanelli:2006,Cea:2007} it has been suggested that, allowing the large-scale spatial geometry of our universe to be plane-symmetric with eccentricity at decoupling of order $10^{-2}$, the quadrupole amplitude can be drastically reduced without affecting higher multipoles of the angular power spectrum of the temperature anisotropy.  Indeed, in Ref.~\cite{Campanelli:2007} it has been shown that the quadrupole anomaly can be resolved if the last scattering surface of CMB is an ellipsoid. In particular, if the eccentricity at decoupling is:
\begin{equation}
\label{Eq.I1} 
e_{\rm dec} \simeq  0.64 \times 10^{-2},
\end{equation}
then the quadrupole amplitude can be drastically reduced without affecting higher multipoles of the angular power spectrum of the temperature anisotropy. 
As discussed in Ref.~\cite{Campanelli:2007},  the anisotropic expansion described by a plane-symmetric metric can be generated by cosmological magnetic fields or topological defects, such as cosmic domain walls or cosmic strings. In fact, topological cosmic defects are relic structures that are predicted to be produced in the course of symmetry breaking in the hot, early universe. Nevertheless, we believe that the most interesting and intriguing possibility is plane-symmetric geometry induced by cosmological magnetic fields for magnetic fields have been already observed in the
universe up to cosmological scales. Remarkably, the estimate of $e_{\rm dec}$, Eq.~(\ref{Eq.I1}), gives for the strength of the cosmic magnetic field:
\begin{equation}
\label{Eq.I2}
B_0 \; \simeq \; 4.6 \times 10^{-9} \;  Gauss  \; ,
\end{equation}
which agrees with the limits arising from primordial nucleosynthesis and large scale structure formation.  \\
\indent
In addition,  in Ref.~\cite{Campanelli:2007}  the direction $(b,l)$ of the axis of symmetry were   constrained to:
\begin{equation}
\label{Eq.I3}
b \; \simeq \; 50^{\circ}\;  - \;  54^{\circ}, \; \;\; 40^{\circ}
\; \lesssim \;  l  \; \lesssim \; 140^{\circ}  \; \; \; or \;  \;  \;  240^{\circ} 
\; \lesssim \; l \; \lesssim \; 310^{\circ} \; ,
\end{equation}
where $b$ and $l$ are the galactic latitude and the galactic longitude, respectively. It turns out that these constraints are in fair agreement with recent statistical
analyses of the cleaned CMB temperature fluctuation maps of the three-year WMAP data obtained by using an improved internal linear combination method as Galactic foreground subtraction technique~\cite{Hinshaw:2007,Oliveira:2006,Park:2007}. Finally, we find amusing that there are already independent indications of a symmetry axis in the large-scale geometry of the universe, coming from the  analysis of polarization of electromagnetic radiation propagating over cosmological distances~\cite{Nodland:1997,Hutsemekers:2005}. \\
\indent
It is known since long time~\cite{Rees:1968, Negroponte:1980,Basko:1980} that anisotropic cosmological models could result in large scale polarization of the cosmic microwave background radiation. Thus, polarization measurements could provide a unique signature of cosmological anisotropy. Indeed, remarkably, the  first year results from Wilkinson Microwave Anisotropy Probe~\cite{1-WMAP} reported statistically significant correlations between the  cosmic microwave background temperature and polarization. The power on small angular scales agrees with the signal expected in models based solely on the temperature spectrum (for a recent review, see Refs.~\cite{Hu:2002,Dodelson:2003}). On the other hand, on large angular scales the detected signal was well in excess of the expected level. This signal on large angular scales has been interpreted as the signature of early reionization. The measured correlation between polarization and temperature yields an electron optical depth to the cosmic microwave background  surface of last scattering of $\tau \simeq 0.17$, corresponding to an instantaneous reionization epoch $z_{re} \simeq 17$. However,  observations of quasars discovered by the Sloan Digital Sky Survey in 2001 at redshifts slightly higher than $z \simeq 6$  do show a Gunn-Peterson trough, indicating that the universe was mostly neutral at redshifts $z \gtrsim 10$. \\
\indent
Since the first Wilkinson Microwave Anisotropy Probe detection of $\tau$, the physics of reionization has been subject to extensive studies~\cite{Cen:2003}-\cite{Mellema:2006}.  However, new results from five-year WMAP data~\cite{Komatsu:2009} indicate that the first year result was an overestimate, and that reionization occurred later,  $z_{re}  =  10.0 \pm 1.4$ and  $\tau = 0.084 \pm 0.016$ . This result is more consistent with the quasar data, although some tension still remains. \\
\indent
In this paper we suggest that the  polarization at large scales detected by the five year Wilkinson Microwave Anisotropy Probe
could be accounted for as  polarization of the cosmic microwave background induced by  the anisotropy of the  spatial geometry  of our universe with  an eccentricity at decoupling given by Eq.~(\ref{Eq.I1}). \\
\indent
The WMAP five-year full-sky maps of the polarization detected at large scales in the foreground corrected maps an average E-mode polarization power~\cite{P-5-WMAP} :
\begin{equation}
\label{Eq.1}
\frac{l(l+1)}{2 \pi} \; C^{EE}_{l= 2} \; = \; 0.150 \; ^{ + 0.427}_{- 0.125} \;  (\mu ^{\circ}K)^2 \; ,
\end{equation}
where the error includes the cosmic variance.  From Eq.~(\ref{Eq.1}) we find:
\begin{equation}
\label{Eq.2}
\frac{(\Delta T)_{pol}}{ T_0} \; =  \; 0.145 \;  ^{+ 0.207}_{- 0,059}  \; 10^{-6} \; ,
\end{equation}
where $T_0 \simeq 2.73 \; ^{\circ}K$. \\
\indent
To evaluate the polarization of the cosmic background radiation induced by eccentricity of the universe we shall follow Ref.~\cite{Cea:2007}. We assume that the photon distribution function $f(\vec{x},t)$ is an isotropically radiating blackbody at a sufficiently early epoch. The subsequent evolution of  $f(\vec{x},t)$ is determined by the Boltzmann equation~\cite{Dodelson:2003}:
\begin{equation}
\label{Eq.3}
\frac{df}{dt} \; = \; C[f] \; , 
\end{equation}
where $ C[f]$ takes care of Thomson scatterings between matter and radiation.  We are interested in the effects of the ellipticity on the Boltzmann equation. For small ellipticity we have~\footnote{Note that we are using natural units where  $\hbar \; = \; c \; =  \; k_B \; =1$.}
:
\begin{equation}
\label{Eq.4}
ds^2 = - dt^2 + a^2(t) (\delta_{ij} + h_{ij}) \, dx^i dx^j \; ,
\end{equation}
where $h_{ij}$ is a metric perturbation which takes on the form:
\begin{equation}
\label{Eq.5}
h_{ij} = -e^2 \delta_{i3} \delta_{j3} \; \; , \; \;e = \sqrt{1 - (b/a)^2} \; .
\end{equation}
Using Eq.~(\ref{Eq.4}) we find:
\begin{equation}
\label{Eq.6}
\frac{df}{dt} = \frac{\partial f}{\partial t} + \frac{\hat{p}^k}{a(t)} \left [1 - \frac{1}{2} h_{ij}  \hat{p}^i  \hat{p}^j \right ]
 \frac{\partial f}{\partial x^k} - p \frac{\partial f}{\partial p} \left [  H(t) +  \frac{1}{2} \dot{h}_{ij} \hat{p}^i  \hat{p}^j \right ] \; ,
\end{equation}
where $H = \dot{a}/a$ is the Hubble rate and $p$ is the photon momentum. To go further we expand the photon distribution about its zero-order Bose-Einstein value:
\begin{equation}
\label{Eq.7}
f_0(p,t) =  \frac{1}{e^{\frac{p}{T(t)}}-1} \; .
\end{equation}
Writing:
\begin{equation}
\label{Eq.8}
f(x,p,\hat{p}^i,t) = f_0(p,t) \left [ 1 + f_1(x,p,\hat{p}^i,t) \right ] \; ,
\end{equation}
we find for the perturbed Boltzmann equation:
\begin{equation}
\label{Eq.9}
\frac{\partial f_1}{\partial t} + \frac{\hat{p}^i}{a(t)} \frac{\partial f_1}{\partial x^i} - \left [  H(t) \,   -  \,  \frac{1}{2} \, \dot{h}_{ij} \right ]  \, 
\hat{p}^i  \hat{p}^j = \frac{1}{f_0} C[f] \; ,
\end{equation}
where we used $\frac{\partial f_0}{\partial \ln p} \simeq -1$ valid in the Rayleigh-Jeans region. \\
\indent
To determine the polarization of the cosmic microwave background we need the polarized distribution function which, in general, is represented by a column vector whose components are the four Stokes parameters~\cite{Chandrasekhar:1960}. In fact, due to the axial symmetry only two Stokes parameters need to be considered, namely the two intensities of radiation with electric vectors  in the plane containing $\vec{p}$ and $\hat{x}^3 \equiv \vec{n}$  and perpendicular to this plane respectively. As a consequence,  we may write:
\begin{equation}
\label{Eq.10}
f(x,p,\hat{p}^i,t) = f_0(p,t) \left[ \pmatrix{1\cr 1\cr} + f_1  \right] \; \; , \; \; f_1 \; = \; \pmatrix{{\xi_1(x,p,\hat{p}^i,t)}\cr  \xi_2(x,p,\hat{p}^i,t)\cr} \; .
\end{equation}
Using Eq.~(\ref{Eq.5}) and defining $\mu \, = \, \cos \theta_{\vec{p}  \vec{n}}$, we get from  Eq.~(\ref{Eq.9}):
\begin{eqnarray}
\label{Eq.11}
&&  \frac{\partial f_1(x,\mu,t)}{\partial t} + \frac{\hat{p}^i}{a(t)} \frac{\partial f_1(x,\mu,t)}{\partial x^i} \;
 = \; \frac{1}{2} \;  \left [ \frac{d}{d \, t} \, e^2(t) \right ] \;  \mu^2  \;  \pmatrix{1\cr 1\cr} \nonumber \\
&& - \sigma_T n_e \left [  f_1(x,\mu,t)   - \frac{3}{8} \int_{-1}^1
\pmatrix{2(1-\mu^2)(1-\mu^{\prime 2})+\mu^2 \mu^{\prime 2}&\mu^2\cr
\mu^{\prime 2}&1\cr} \,  f_1(x,\mu',t) d\mu' \right] 
\end{eqnarray}
where $\sigma_T$ is the Thomson cross section and $ n_e(t)$  the electron number density. Introducing the conformal time:
\begin{equation}
\label{Eq.12}
\eta(t) \; =  \;  \int_{0}^t \frac{dt'}{a(t')} \; \; ,
\end{equation}
we rewrite Eq.~(\ref{Eq.11}) as:
\begin{eqnarray}
\label{Eq.13}
 && \frac{\partial f_1(\mu,\eta)}{\partial \eta} + \hat{p}^i  \frac{\partial f_1(\mu,\eta)}{\partial x^i}  \; 
= \; \frac{1}{2} \;  \frac{d }{d \eta}e^2(\eta)  \; (\mu^2 - \frac{1}{3})   \pmatrix{1\cr 1\cr}  \nonumber \\
&&  - \sigma_T n_e  a(\eta) \left [  f_1(\mu,t)   - \frac{3}{8} \int_{-1}^1
\pmatrix{2(1-\mu^2)(1-\mu^{\prime 2})+\mu^2 \mu^{\prime 2}&\mu^2\cr
\mu^{\prime 2}&1\cr} \,  f_1(\mu',\eta) d\mu' \right]
\end{eqnarray}
with a suitable overall normalization of the blackbody intensity.  Since we are interested in the long wavelenght limit, we may neglect the spatial derivative term in Eq.~(\ref{Eq.13}).  In this case, the general solutions of Eq.~(\ref{Eq.13}) can be written as~\cite{Basko:1980}:
\begin{equation}
\label{Eq.14}
 f_1(x,\mu,\eta) =   \theta_a(\eta) (\mu^2 - \frac{1}{3})   \pmatrix{1\cr 1\cr} +   \theta_p(\eta) (1 - \mu^2 ) 
   \pmatrix{1\cr -1\cr} \; .
\end{equation}
From Eq.~(\ref{Eq.14}) it is evident that  $ \theta_a(\eta)$ measures the degree of anisotropy, while $\theta_p(\eta)$ gives the polarization of the primordial radiation. Following Ref.~\cite{Basko:1980} we find:
\begin{equation}
\label{Eq.15}
  \theta_a(\eta) = \frac{1}{7}  \int_{0}^{\eta} \Delta H(\eta') \left[ 6 e^{-\tau(\eta,\eta')}   + e^{- \frac{3}{10} \tau(\eta,\eta')}  \right] d\eta' \; \; ,
\end{equation}
\begin{equation}
\label{Eq.16}
 \theta_p(\eta) = \frac{1}{7}  \int_{0}^{\eta} \Delta H(\eta') \left[ e^{-\tau(\eta,\eta')}   -  e^{- \frac{3}{10} \tau(\eta,\eta')}  \right] d\eta' 
 \; \; ,
\end{equation}
where we introduce the optical depth:
\begin{equation}
\label{Eq.17}
 \tau(\eta,\eta') =   \int_{\eta'}^{\eta} \sigma_T \, n_e \, a(\eta'') \, d\eta'' \; \; ,
\end{equation}
and the cosmic shear~\cite{Negroponte:1980}:
\begin{equation}
\label{Eq.18}
\Delta H(\eta) \;  = \;  \frac{1}{2} \; \frac{d }{d \eta}e^2(\eta)  \; \; .
\end{equation}
Equation~(\ref{Eq.16}) shows that the polarization of the cosmic microwave background (without reionization) at the present time is essentially that produced around the time of recombination, since much later the free electron density is negligible ( $\tau \simeq 0$), while much earlier the optical depth is very large. Then, the present polarization  is the result of Thomson scattering around the time of decoupling of matter and radiation, which occurs after the free electron density starts to drop significantly~\cite{Peebles:1993}.  On the other hand, it is evident from Eq.~(\ref{Eq.15}) that the main contribution to the temperature anisotropy $\theta_a$ comes from conformal time $\eta \; \gtrsim \; \eta_d$, where  $\eta_d$ the conformal time around which decoupling occurs. Since soon after decoupling the optical depth vanishes, we get:
\begin{equation}
\label{Eq.15b}
  \theta_a(\eta) \; \simeq \;    \int_{\eta_d}^{\eta} \Delta H(\eta')  d\eta' \; \; .
\end{equation}
Using Eq.~(\ref{Eq.18}) we easily evaluate the anisotropy at the present time $\eta_0$:
\begin{equation}
\label{Eq.15c}
 \theta_a(\eta_0) \; \simeq \;    \int_{\eta_d}^{\eta_0}  \Delta H(\eta')  d\eta' \; \; = \;  \frac{1}{2}\;  \int_{t_d}^{t_0}   \frac{d }{d t'}e^2(t') 
 \;  d t'  \; = \; -\; \frac{1}{2} \;  e^2_{\rm dec} \; ,
\end{equation}
where we used $e(t_0) \, = 0$. Obviously, our result Eq.~(\ref{Eq.15c}) agrees with  the temperature anisotropy evaluated with the Sachs-Wolfe effect~\cite{Campanelli:2006,Cea:2007,Campanelli:2007}. \\
\indent
To evaluate $\theta_p$ we need to determine the   cosmic shear Eq.~(\ref{Eq.18}) at decoupling since, as we said before, the polarization of the cosmic microwave background (without reionization) at the present time is essentially that produced around the time of recombination. To this end we shall restrict to the most interesting case of plane-symmetric geometry induced by cosmological magnetic fields. In this case, from the results in Ref.~\cite{Campanelli:2007} we get:
\begin{equation}
\label{Eq.16b}
e^2(t) = 8 \Omega_B^{(0)} (1 - 3 a^{-1} + 2 a^{-3/2}) \; ,
\end{equation}
where $\Omega_B^{(0)} = \rho_B(t_0)/\rho_{\rm cr}^{(0)}$, and $\rho_{\rm cr}^{(0)} = 3 H_0^2/8 \pi G$ is the actual critical
energy density.  Note  that we are adopting the  normalization such that $a(t_0) = 1$ and $e(t_0) = 0$. Since in the matter-dominated era  $a(t) \propto t^{2/3}$, so that $H = \frac{2}{3t}$, we may write near decouplig:
\begin{equation}
\label{Eq.16c}
  \frac{1}{2} \; \frac{d }{d t}e^2(t) \; \simeq \; - \; \frac{3}{4} \;  e^2(t) \; H(t) \; .
\end{equation}
Inserting Eqs.~(\ref{Eq.18})  and (\ref{Eq.16c})  into Eq.~(\ref{Eq.16})  we obtain:  
\begin{equation}
\label{Eq.19}
 \theta_p(\eta) \;  \simeq  \;   - \frac{3}{28}  \int_{0}^{\eta} e^2(\eta') H(\eta') \left[ e^{-\tau(\eta,\eta')}   -  e^{- \frac{3}{10} \tau(\eta,\eta')}  \right] d\eta'  \; \; .
\end{equation}
We, now, define $\tau(\eta) =  \tau(\eta_0, \eta)$, so that $\tau(\eta', \eta) = \tau(\eta') - \tau(\eta)$. After that, changing the integration variable and using~\cite{Harari:1993}:
\begin{equation}
\label{Eq.20}
\frac{d}{d \eta} \tau(\eta) \; \simeq \; - \frac{\tau(\eta)}{\Delta  \eta_d} \; \; ,
\end{equation}
where $\Delta  \eta_d$ is the conformal time  duration of the  decoupling process, we get:
\begin{equation}
\label{Eq.21}
\theta_p(\eta) \simeq  - \frac{3}{28} \, e^2(\eta_d) \, H(\eta_d) \, \Delta \eta_d \, e^{\tau(\eta)} \, \int_{0}^{\infty} \frac{dz}{\tau(\eta)+z} \, e^{-z} \left[ 1  -  e^{ \frac{7}{10} z } \right] \; .
\end{equation}
So that we have:
\begin{equation}
\label{Eq.22}
\theta_p(\eta_0) \simeq   \frac{3}{28} \ln{\frac{10}{3}} \, e^2(\eta_d) \, H(\eta_d) \, \Delta \eta_d  \; ,
\end{equation}
which can be rewritten as:
\begin{equation}
\label{Eq.23}
\theta_p(\eta_0) \;  \simeq  \;  \frac{3}{28} \ln{\frac{10}{3}} \, e^2(t_d) \,\frac{ \Delta z_d}{1 + z_d} \; .
\end{equation}
To evaluate $ \theta_p(\eta_0)$ we may assume that $ \Delta z_d \sim 10^{2}$, $z_d \sim 10^{3}$ which together with Eq.~(\ref{Eq.I1}) leads to our final result:
\begin{equation}
\label{Eq.25}
\theta_p = \frac{(\Delta T)_{pol}}{ T_0}  \simeq  0.53 \, 10^{-6} \; \; .
\end{equation}
If we consider the mass density parameter $\Omega_m$, then we should multiply Eq.~(\ref{Eq.25}) by a factor $1/\sqrt{\Omega_m} \; \simeq \; 2.0$.
Thus, we find for the average large scale polarization:
\begin{equation}
\label{Eq.26}
 \frac{(\Delta T)_{pol}}{ T_0}  =  (0.5 \;  -  \; 1.0) \;  10^{-6} \; \; .
\end{equation}
Note that Eq.~(\ref{Eq.23})  gives the average  large scale  polarization of the cosmic microwave background induced by  the anisotropy of the  spatial geometry  of the  universe. On the other hand, it is evident from our discussion that for wavelenghts comparable or smaller than the width of the last scattering surface, the polarization should fall off very rapidly. Indeed, the polarization signal should be confined up to multipoles $ l_{max}$ such that $\frac{1}{l_{max}}  \sim \frac{ \Delta z_d}{ z_d}  \sim 10^{-1}$.  More precise statements can be obtained by solving numerically the radiative transfer equation for the cosmic microwave background including polarization in anisotropic universes. Indeed, recently Pontzen and  Challinor~\cite{Pontzen:2007}   have 
derived the  radiative transfer equation  in the nearly-Friedmann-Robertson-Walker limit of homogeneous, but anisotropic, universes classified via their Bianchi type. These authors argued that  the  polarization signal is mostly confined to multipoles $l \lesssim 10$.

\indent
In conclusion, we see that  the proposal of a small  anisotropy in  the  large-scale spatial geometry  of our universe~\cite{Campanelli:2006,Cea:2007,Campanelli:2007}  could result in a drastic reduction in the quadrupole anisotropy without affecting higher multipoles of the angular power spectrum of the temperature anisotropy, and gives rise to a sizeable large-scale  polarization of the cosmic microwave background. Indeed, the WMAP five-year full-sky maps of the polarization detected at large scales  an average E-mode polarization power~\cite{P-5-WMAP}.  However,  Eq.~(\ref{Eq.26})  shows that the large scale  polarization of the cosmic microwave background induced by  the anisotropy of the  universe geometry  seems to exceed  the average level of  polarization detected by the Wilkinson Microwave Anisotropy Probe, Eq.~(\ref{Eq.2}), by  at least a factor of $2$.  We feel, however, that the instrumental limitations of the Wilkinson Microwave Anisotropy Probe could have result in an overestimation of the foreground polarization signal. In fact, a careful characterization  of foreground polarization is certainly crucial for polarization measurements. Given that Planck has  broader frequency coverage to subtract foregrounds, we expect that the fortcoming polarization data at large scales from Planck will be able to definitely corroborate or reject the proposal for  anisotropic universe.
\end{document}